\documentclass[preprint, 14pt]{elsarticle}
\usepackage[top=0.65in,left=1.0in,footskip=0.75in]{geometry}

\usepackage{amssymb}
\usepackage[cmex10]{amsmath}
\usepackage{color,xcolor,soul,framed} 
\usepackage{graphicx}
\usepackage[font=normalfont]{caption} 
\usepackage{multirow}
\biboptions{numbers,sort&compress}

\definecolor{turquoise}{rgb}{0, 0, 0}
\definecolor{turquoiseee}{rgb}{.0,.0,.0}
\definecolor{red_remove}{rgb}{1.0,0.0,0.0}

\usepackage{hyperref}
\hypersetup{
    colorlinks=true,
    citecolor=blue,
    linkcolor=red,
    filecolor=magenta,      
    urlcolor=blue,
    pdftitle={Overleaf Example},}
\newcommand{\vv}{\checkmark} 

\journal{Measurements}
\begin{document}

\begin{frontmatter}
\title{Data-Driven Denoising of Stationary Accelerometer Signals}
\author[inst1]{Daniel Engelsman}

\affiliation[inst1]{organization={The Hatter Department of Marine Technologies, Charney School of Marine Sciences, University of Haifa},  city={Haifa}, country={Israel}}

\author[inst1]{Itzik Klein}


\begin{abstract}
Modern navigation solutions are largely dependent on the performances of the standalone inertial sensors, especially at times when no external sources are available. During these outages, the inertial navigation solution is likely to degrade over time due to instrumental noises sources, particularly when using consumer low-cost inertial sensors. Conventionally, model-based estimation algorithms are employed to reduce noise levels and enhance meaningful information, thus improving the navigation solution directly. However, guaranteeing their optimality often proves to be challenging as sensors performance differ in manufacturing quality, process noise modeling, and calibration precision. In the literature, most inertial denoising models are model-based when recently several data-driven approaches were suggested primarily for gyroscope measurements denoising. Data-driven approaches for accelerometer denoising task are more challenging due to the unknown gravity projection on the accelerometer axes. To fill this gap, we propose several learning-based approaches and compare their performances with prominent denoising algorithms, in terms of pure noise removal, followed by stationary coarse alignment procedure. Based on the empirical benchmarking, we show that the learning-based models outperform traditional signal processing filtering in terms of pure inertial signal reconstruction. Moreover, they are shown to improve angular errors by one order of magnitude, given a navigation-related task.
\end{abstract}




\end{frontmatter}


\section{Introduction} \label{sec:intro}
Inertial Navigation System (INS) is one of the most commonly used navigation systems.  It consists of an inertial measurement units (IMU) with three orthogonal gyroscopes and accelerometers to determine the platform position, velocity, and orientation \cite{Jekeli2000, Groves2013}. With recent technological rise of micro-electro-mechanical-systems (MEMS) inertial sensors, their integration became very common in many applications, platforms, and environments (air, ground, sea). Their popularity can be explained due to their small size, high cost-effective, low power consumption, and low-cost prices. However, when external position/velocity update source is not available, the INS navigation solution becomes solely dependent on the inertial sensor performance and its error regime. Integration of noisy measurements propagates into the navigation solution, resulting in a solution drift \cite{Titterton2004}. This problem worsens when low cost MEMS-IMU are used, due to their characteristic noise and inherent bias \cite{el2007analysis}. \\
To reduce the influence of the sensor error terms, its readings are passed through a signal processing algorithm for signal denoising followed by a calibration procedure to estimate the bias, scale-factor, and misalignment error terms. Generally, the calibration procedure has two steps: 1) estimating the deterministic parts of the sensor errors and 2) using estimation techniques to remove the residuals from the first step and the stochastic parts of the error terms. For this task, in stationary conditions, the Kalman filter can be used, exploiting the fact that a stationary sensor allows zero velocity updates. However, when a thorough calibration process cannot be taken, conventional filtering cannot help mitigating these errors. This is where the powerful capabilities of learning approaches come into the picture, enabling denoising based on patterns extracted from data.
\\
According to our literature review, among all data-driven denoising works, only one work addresses accelerometer denoising, as the rest perform gyroscopes denoising. It is not surprising in light of the instrumental differences, as same physical scenario is projected inherently different over each sensor. For example, while stationary, MEMS gyroscopes (insensitive to Earth's rotation rate) are expected to output a three-dimensional zero vector $\| \boldsymbol{\omega} \| \approx 0$, regardless body orientation (in practice we get sensor noise). 
Contrarily, MEMS accelerometers sense Earth gravitation depending on a given orientation, such that their axes being regularly subjected to a larger norm $\| \boldsymbol{f} \| \approx \text{g}$. The constant presence of non-zero amplitudes, wider dynamic ranges and larger noise densities \cite{dadafshar2014accelerometer}, are ultimately reflected in the scale and variability of the samples. Additionally, these differences also propagate into the calibration process. While gyroscopes calibration is self-contained as output offset can be simply subtracted from zero, accelerometers require auxiliary attitude information to ensure parallelism to the gravity axis. When data-driven models are required to generalize over these patterns, distributions of specific force measurements exhibit significantly higher levels of variance and bias, making accelerometers denoising much more challenging, thus still unexplored.
\\
To fill this gap, this work presents an innovative denoising approach, exploiting the powerful generalizability of learning based approaches and their robustness to noisy patterns. The contributions of this paper can be summarized as follows:
\begin{enumerate}
\item We propose a learning-based approach for denoising stationary accelerometers, outperforming conventional signal processing-based denoising techniques.
\item We assess the denoising contribution with a navigation-related metric: stationary coarse alignment (SCA), showing an improvement of one order of magnitude.
\item We give a comprehensive literature review of former related works to target our contribution.
\end{enumerate}

The rest of the paper is organized as follows: Section \ref{sec:LR} gives an up-to-date literature review of MEMS-IMU denoising methods, Section \ref{sec:CA} describes auxiliary methods used to assess our strategy, and section \ref{sec:PA} presents our proposed solution. Section \ref{sec:DnM} elaborates on the data preparation process, and section \ref{sec:AandR} shows analysis of the results. Section \ref{sec:limit} describes the limitations of the study, and finally, Section \ref{sec:CN} gives conclusions. 

\section{Literature Review}\label{sec:LR}
Study of inertial sensor denoising techniques dates back to the late 90s, and can be inclusively divided into conventional signal processing approaches and to recent learning-based approaches. Followed is an in-depth literature review to conveniently map the problem domain. Some works evaluate their proposed denoisers in terms of reconstruction of the original IMU output, while others examine its contribution as a prefilter upon a wider endgame output (e.g. Euler angles, navigation states). 
\begin{table}[!h]
\caption{Summary of SP-based denoisers} \label{t:stats}
\renewcommand{\arraystretch}{1.05}
\centering
\begin{tabular}{c|c|cc|cc|}
Algorithm   & Ref. & Gyro & Acc. & Stationary & Dynamic \\ [1mm] \hline 
Moving average & \cite{gonzalez2018statistical} & \vv &            & \vv & \vv \\[.5mm] \hline 
\multirow{3}{*}{ARMA} & \cite{waegli2010noise, diao2013analysis, yong2015research, tu2020arma, abbasi2022memory}  & \vv   &     & \vv   &   \\ 
        & \cite{nassar2004modeling, nassar2005accurate, wang2018time} & \vv       & \vv        & \vv   &   \\
        & \cite{yuan2010research}   & \vv        & \vv        &       & \vv \\\hline
\multirow{3}{*}{EMD}    & \cite{gan2014emd} & \vv   & \vv        &  \vv       & \vv \\
        & \cite{liu2019gyroscope, wang2021research}   & \vv        &           & \vv &  \\
        & \cite{shen2016noise, guo2018hybrid, liu2020denoising} & \vv        &           & \vv & \vv \\ \hline
\multirow{3}{*}{Savitsky-Golay} & \cite{li2013noise}        &     & \vv & \vv  &     \\
      & \cite{nirmal2016noise, karaim2019low} & \vv & \vv &  \vv & \vv \\
      & \cite{he2019noise}      & \vv &  &  \vv & \vv \\ \hline
\multirow{4}{*}{Wavelets}                    & \cite{kang2010wavelet, kang2011improvement}&\vv &  \vv  & & \vv \\
                    & \cite{el2004wavelet}      & \vv & \vv & \vv &  \\
                    & \cite{liu2007mems, li2014improved, song2019mems}        & \vv &  & \vv  &  \\ 
                    & \cite{qu2009adaptive, yuan2015improved, el2018utilization, Ali_2021}       & \vv &  &      & \vv \\\hline
\end{tabular}
\label{t:sp}
\end{table}
Table \ref{t:sp} presents conventional signal processing (SP-based) approaches whose analysis and synthesis is based on the signal structure and nature, enabling to detect components of interest. 
Moving average (MA) techniques can be used as efficient smoothing filter, based on errors (residuals) from previous forecasts \cite{gonzalez2018statistical}. However, determining its optimal window size is mostly heuristic, depending largely on the characteristics of a given dataset. Other works elaborated this by combining a weighted regression term over the lagged values, namely auto-regressive moving-average (ARMA) \cite{waegli2010noise, diao2013analysis, yong2015research, tu2020arma, abbasi2022memory, nassar2004modeling, nassar2005accurate, wang2018time, yuan2010research}. Empirical mode decomposition (EMD) is a time frequency analysis which decomposes multicomponent signals into a finite number of Intrinsic Mode Functions (namely, its building blocks), thus emphasizing local characteristics such that non-stationary and nonlinear signals can be robustly handled \cite{gan2014emd, liu2019gyroscope, wang2021research, shen2016noise, guo2018hybrid, liu2020denoising}. Yet its sifting process is sensitive to aperiodic fluctuations and volatile trends. 
Savitsky-Golay is a smoothing filter that fits an optimal local curve over a moving window size, using low-degree polynomial regression \cite{li2013noise, nirmal2016noise, karaim2019low, he2019noise}. \\
However despite its popularity, its polynomial fitting requires heuristic treatment for elements in the window ends, and suppression of high frequencies tends to underperform due to oversmoothing. And finally, the wavelet-based algorithms \cite{kang2010wavelet, kang2011improvement, el2004wavelet, liu2007mems, li2014improved, song2019mems, qu2009adaptive, yuan2015improved, el2018utilization, Ali_2021}, which are considered as most popular for MEMS-IMU denoising. Using a shifted and scaled window function, time signal is projected into a set of basis functions named wavelets. Conversely, it can be computationally intensive for fine analysis, and its transform exhibits poor shift invariance when high volatility is introduced.
Table \ref{t:ml} presents a group of learning-based algorithms, which gained much popularity in recent years in a growing number of fields \cite{khanafer2020applied}, including inertial sensors and autonomous navigation. Their advantage lies in their ability to identify complex patterns by learning high-level features of the data, thus diminishing the need in domain expertise.
\begin{table}[ht]
\caption{Summary of learning-based denoisers}
\renewcommand{\arraystretch}{1.1}
\centering
\begin{tabular}{c|c|cc|cc|}        
Algorithm   & Ref. & Gyro & Acc. & Stationary & Dynamic \\ [1mm] \hline 
Linear regression & \cite{gonzalez2019time} & \vv & \vv & \vv & \vv \\ \hline 
CNN     & \cite{brossard2020denoising, huang2022mems} & \vv   &    & \vv   & \vv  \\\hline 
RNN     & \cite{jiang2018mems, jiang2018performance, ruoyu2019modeling} & \vv   &    & \vv   &   \\\hline
\multirow{2}{*}{LSTM}    & \cite{zhu2019mems, zhu2021combined} & \vv   &    & \vv   &   \\
        & \cite{han2021hybrid} & \vv   &    & \vv & \vv \\
\hline
GRU     & \cite{jiang2019mixed} & \vv   &    & \vv   &   \\ \hline
\end{tabular}
\label{t:ml}
\end{table}

Gonzalez proposed a multiple linear regression (MLR) model which uses several vector-valued variables to predict the outcome of a dependent variable \cite{gonzalez2019time}. Bossard proposed a convolutional neural network (CNN) which computes gyro corrections for the undesirable noise, before integrated into orientation increments \cite{brossard2020denoising, huang2022mems}. The rest of the references utilized a common extension to feedforward neural networks called recurrent neural networks (RNN), which were designed to handle variable-length sequential signals. Unlike CNNs, which excel at finding spatial relations over grid-like topology, RNNs have a feedback element which enables forward and backwards connections such that complex dynamic relationships over distant time steps are better captured \cite{jiang2018mems, jiang2018performance, ruoyu2019modeling}. \\
However, despite its ability to process temporal information of any length, it is limited with long-term dependencies due to the exponential decay of the loss function, namely vanishing gradient problem. To that end, an improved versions were introduced, where additional control units (gates) were added to allow better flow of the gradient and to maintain memory over long time periods, namely long short term memory (LSTM) \cite{zhu2019mems, zhu2021combined, han2021hybrid} and gated recurrent unit (GRU) \cite{jiang2019mixed}. As already discussed above, it can be seen that the majority of the works (75$\%$) address gyroscopes denoising, especially in the learning-based approaches.

\section{Problem Formulation}\label{sec:CA}
After describing the big picture, this section presents two functional methods used for benchmarking (\textit{A.}) and evaluating (\textit{B.}) the validity of our proposed approaches. 

\subsection{Signal Processing Denoising Approaches} \label{subsec:sp}
Conventional SP-based denoisers perform noise reduction by either spatial smoothing, local regression, or by imposing spectral constraints to filter out unwanted frequencies. Following are three reference methods, used for comparison with the data-driven approaches. Their optimality was obtained by brute-force search over the training set.\\

\subsubsection{Moving Average (MA)} suppresses unstable signal noises by averaging measurements inside a rolling window \cite{gonzalez2018statistical}
\begin{align}
\hat{\boldsymbol{x}}_{\textbf{\textit{MA}},i} = \frac{1}{T} \sum_{t = 0}^{T-1} \boldsymbol{x}_{i+t-T} \quad \forall \ i \geq T ,
\end{align}
where $\boldsymbol{x}_{i}$ is a noisy sample and $T$ is the window size.

\subsubsection{Savitsky-Golay (SG)} allows denoising by deriving observations directly from time domain, thus avoiding spectral decomposition \cite{li2013noise, nirmal2016noise, karaim2019low, he2019noise}. By fitting successive sets of adjacent points with a low-degree polynomial, followed by least-squares regression, local noise is smoothed out. Noisy signal $\boldsymbol{x}_{i}$ is replaced with a set of $m$ convolution coefficients $C_t$,
\begin{align}
\hat{\boldsymbol{x}}_{\textbf{\textit{SG}},i} = \sum_{t=-t_s}^{t_s} C_t  \boldsymbol{x}_{i+t} \hspace{1cm}  \tfrac{m+1}{2} \leq t \leq t_f - \tfrac{m-1}{2} \ ,
\end{align}
where $t_f$ is the duration of data point $x_t$ and $t_s = \frac{m-1}{2}$ denotes the window margins. 

\subsubsection{Discrete wavelet transform (DWT)} the most common denoising technique, where input signal $\boldsymbol{x}_i$ is represented in both time and frequency domains, by decomposing it into a set of basis functions \cite{kang2010wavelet, kang2011improvement, el2004wavelet, liu2007mems, li2014improved, song2019mems, qu2009adaptive, yuan2015improved, el2018utilization, Ali_2021}. Here, a Daubechies (db4) mother wavelet function ($\psi$) is used with different scaling ($a=2$) and shifting ($b$) parameters, providing a progressively finer outputs, given by coefficients matrix $\Psi$
\begin{align}
\Psi_i[b,a^j] &= \sum_{t=0}^{N-1} \boldsymbol{x}_i[t] \frac{1}{\sqrt{a^j}} \psi_{j} \left( \frac{t-b}{a^j} \right).
\end{align}
Since small valued coefficients are dominated by noise, hard thresholding $\text{T}_{\text{Hard}}$ is used to remove them, thus preserving only meaningful information. Then, an inverse transform is applied on the thresholded wavelet coefficients, to reconstruct the denoised matrix back to time domain signal as given by
\begin{align}
\hat{\boldsymbol{x}}_{\textbf{\textit{DWT}},i} = \Big( \text{T}_{\text{Hard}} (\Psi_{i} ) \Big)^{-1} .
\end{align}

\subsection{Inertial Navigation Coarse alignment}
In addition to analyzing our proposed approach denoising capabilities on the accelerometer signals, we examine its influence on a stationary coarse alignment (SCA) procedure. To that end, this section gives a brief introduction to the coarse alignment theory. The attitude (roll and pitch angles) can be determined using accelerometer measurements. 
The orientation of body coordinates frame with respect to navigation frame is given by a transformation matrix, represented by a product of three successive rotations about the z-y-x axes, corresponding to Euler angels: yaw ($\psi$), pitch ($\theta$) and roll ($\phi$), where s and c stand for sine and cosine, respectively.
\begin{equation} \label{eq:T_b_n}
\textbf{T}_{b}^n = \begin{bmatrix}
\text{c}_\theta  \text{c}_\psi  &  \text{s}_\phi \text{s}_\theta \text{c}_\psi - \text{c}_\phi \text{s}_\psi  &   \text{c}_\phi \text{s}_\theta \text{c}_\psi + \text{s}_\phi \text{s}_\psi \\
\text{c}_\theta \text{s}_\psi & \text{s}_\phi \text{s}_\theta \text{s}_\psi +  \text{c}_\phi \text{c}_\psi & \text{c}_\phi \text{s}_\theta \text{s}_\psi - \text{s}_\phi \text{c}_\psi  \\
-\text{s}_\theta & \text{s}_\phi \text{c}_\theta & \text{c}_\phi \text{c}_\theta 
\end{bmatrix}.
\end{equation}

In stationary conditions, accelerations in navigation frame are equal to zero ($\dot{\textbf{v}}^n=\textbf{0}$), thus only the gravity vector \textbf{g}, is projected upon the accelerometer axes \cite{Titterton2004} 
\begin{align} \label{eq:gravity}
\boldsymbol{f}_{ib}^b = 
\begin{bmatrix} 
f_{ib,x}^b \\ f_{ib,y}^b \\ f_{ib,z}^b
\end{bmatrix} = -\textbf{T}_n^b \, \textbf{g}^n  = \begin{bmatrix} 
s_{\theta} \\ - s_{\phi} c_{\theta}  \\ - c_{\phi} c_{\theta} 
\end{bmatrix} \text{g}.
\end{align}
Using analytical coarse alignment, leveling can be computed by the axial components of the specific force vector $\boldsymbol{f}_{ib}^b$
\begin{align}
\phi &= \arctan_2 \Big( -f_{ib,y}^b \, , \, -f_{ib,z}^b \Big) , \label{eq:roll} \\
\theta &= \arctan \Big( \frac{ -f_{ib,x}^b }{ \sqrt{ f_{ib,y}^{b \ 2} + f_{ib,z}^{b \ 2} } } \Big). \label{eq:pitch}
\end{align}
The accuracy of the SCA process affects the whole INS performance, especially in pure inertial navigation. To increase the degraded signal-to-noise ratio (SNR), instrumental errors can be reduced in proportion to a square root factor, by either averaging over $n$ identical sensors ($1/\sqrt{n}$), or by averaging over $T$ time steps ($1/\sqrt{T}$). Thereby, effective stationary denoising can reduce averaging times of the SCA procedure, or improve the alignment accuracy over the same period of time.

\section{Proposed Approach}\label{sec:PA}
\begin{figure}[h]
\begin{center}
\includegraphics[width=0.75\textwidth]{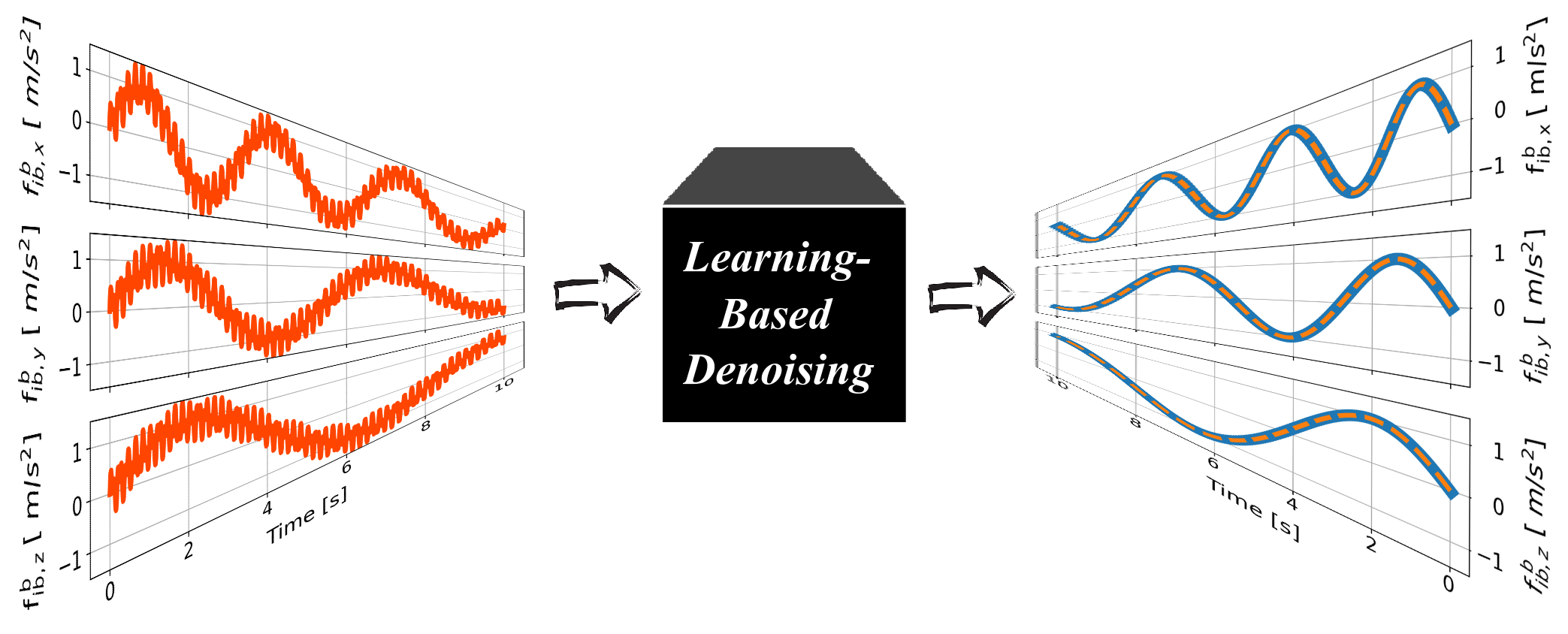}
\caption{Denoising mapping function  $f : \boldsymbol{x} \rightarrow \hat{\boldsymbol{x}}$ }
\label{f:Denoising-General}
\end{center}
\end{figure}

We propose implementing and modifying several learning algorithms for the stationary accelerometer denoising problem, which unlike conventional, model-based, denoising filters, are independent in any analytic error model. Instead, they learn to map noisy readings into their corresponding accurate measurements. Fig.~\ref{f:Denoising-General} presents the proposed approach, where noisy measurements from commercial grade inertial sensors are input to a learning model for the denoising task. The model then outputs denoised accelerometer readings, which are used to calculate an approximation error, with respect to a high-end, accurate inertial sensors readings, as ground-truth (GT). 
The error is first used to optimize the model during the training phase. Then, during testing phase (\ref{sec:AandR}), the well-trained models are evaluated over unseen noisy samples. Unlike SP-based methods, here extracted patterns are associated with target functions, enabling signal denoising by generalization. Similarly to human processing, feedforward neural networks (FNN) excel at pattern recognition, as their layered structure perform hierarchical representations, allowing extraction of spatial features \cite{schmidhuber2015deep}. However given sequential relations, their node-specific weights eventually fail to learn, as lengthy inputs and complex ordering impose an exponentially growing number of parameters. To that end, a recurrent feedback mechanism was proposed \cite{elman1990finding}.
\begin{figure}[h]
\begin{center}
\includegraphics[width=0.55\textwidth]{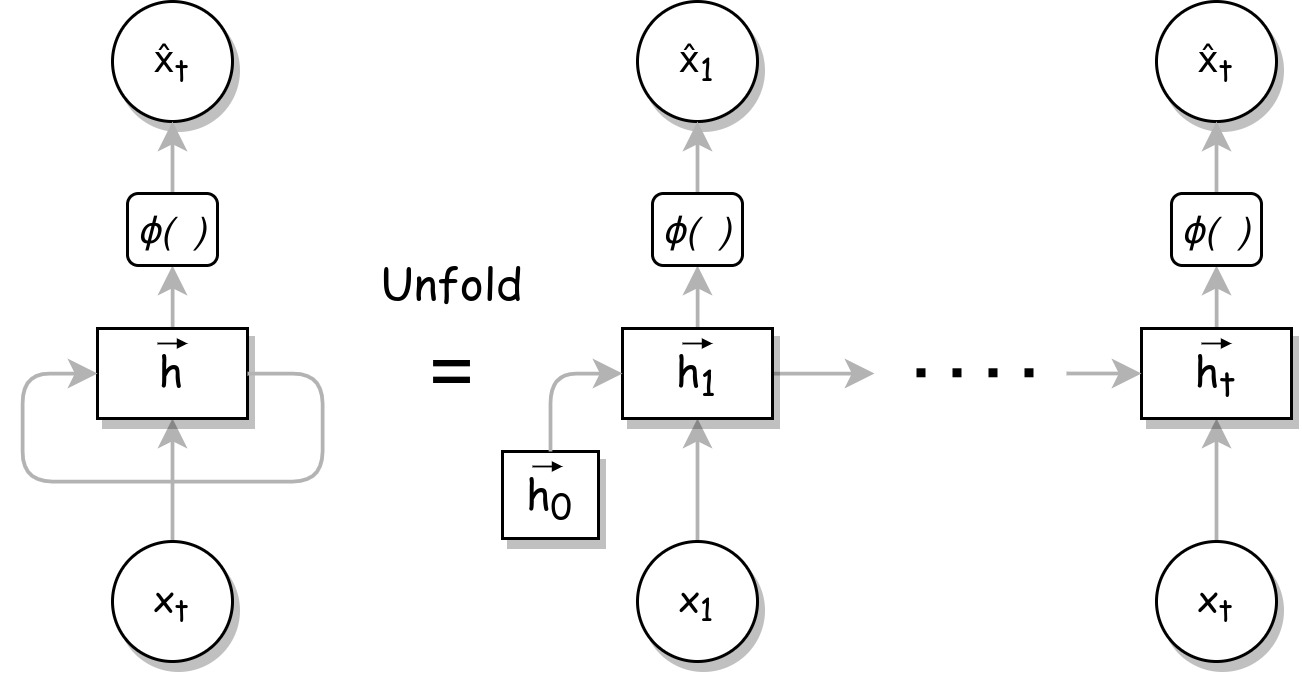}
\caption{One-layer recurrent neural network (RNN)}
\label{f:RNN_basic}
\end{center}
\end{figure}
Three dimensional measurements are processed inside an encoding cell called hidden state, using a weight vector of $m$ learnable parameters, such that $\overrightarrow{\textbf{\textit{h}}} \in \mathbb{R}^{3 \times m}$. Unlike FFN, here same weights are shared across the entire sequence, and current inputs are taken from previous outputs. This way, the model avoids spatial memorization and focuses on intertemporal dependencies. Fig.~\ref{f:RNN_basic} illustrates a sequence-to-sequence mapping, consists of single 3D noisy measurements, estimated (denoised) every time step $ \hat{x}_t \ \in \mathbb{R}^3$. Following, we elaborate on our proposed models: three variations of deep recurrent neural networks (RNN) and one naive machine learning model (kNN). 

\subsection{Unidirectional bi-layer LSTM} 
The architecture of the first model is similarly arranged as shown in Fig.~\ref{f:RNN_basic}, but consists of two (bi) stacked layers. The hidden states are an RNN variant called long-short term memory (LSTM) \cite{hochreiter1997long}, originally developed to handle the decay of the loss function. The forget gate $\textbf{\textit{f}}_t$ decides whether ignoring or adding new (short-term) information to the cell state $\textbf{\textit{C}}_t$ memory. The input gate $\textbf{\textit{i}}_t$, is then used to estimate the relevance of current input to past hidden states. This way, the cell state acts as a global memory unit, where distant dependencies (long-term) are maintained across time
\begin{align}
\textbf{\textit{f}}_t &= \sigma( \textbf{W}_f  [ x_t, \, \overrightarrow{h}_{t-1} ] + \textbf{b}_f  ), \\
\textbf{\textit{i}}_t &= \sigma( \textbf{W}_i \hspace{.7mm}  [ x_t, \, \overrightarrow{h}_{t-1} ] + \textbf{b}_i  ), \\
\tilde{\textbf{\textit{C}}}_t &= \tanh ( \textbf{W}_c  [x_t, \, \overrightarrow{h}_{t-1}] + \textbf{b}_c  ),  \\
\textbf{\textit{C}}_t &= \textbf{\textit{f}}_t \odot \textbf{\textit{C}}_{t-1} + \textbf{\textit{i}}_t \odot \tilde{\textbf{\textit{C}}}_t .
\end{align}
Operator $\odot$ denotes an element-wise multiplication, $\textbf{W}_j$ and $\textbf{b}_j$ are the weights and biases respectively, and the sigmoid function scales encoded vectors as follows $\sigma:\mathbb{R}^m \rightarrow (0, 1)^m$. 
Next, the output gate $\textbf{\textit{o}}_t$ is fused with the updated cell state 
\begin{align}
\textbf{\textit{o}}_t &= \sigma( \textbf{W}_o \hspace{.5mm}  [ x_t, \, \overrightarrow{h}_{t-1} ] + \textbf{b}_o  ) ,
\end{align}
such that the next hidden state is determined as follows:
\begin{align}
\overrightarrow{\textbf{\textit{h}}_t} &= \textbf{\textit{o}}_t \odot \tanh ( \textbf{\textit{C}}_t ) .
\end{align}
Finally, the hidden state signal is taken for state prediction, using a linear activation function $\boldsymbol{\phi}$ (neuron) which scales it back into a continuous and unbounded range, given by
\begin{align}
\hat{\boldsymbol{x}}_t &= \boldsymbol{\phi}( \overrightarrow{\textbf{\textit{h}}_t} ) = \textbf{W}_{\hat{y}}^{\mathsf{T}} \overrightarrow{\textbf{\textit{h}}_t} + \textbf{b}_{\hat{y}} \ \in \ \mathbb{R}^3 .
\end{align}

\subsection{Bi-directional one-layer RNN}
Conventionally, sequential inputs are analyzed along the positive time direction (chronologically), thus limiting the learning with respect to only past states. Using two opposing cells, future contexts can be learned anti-chronologically, compensating the lack of cell state concept. Here, two different hidden layers are allocated to process data both in forward $\overrightarrow{\textbf{\textit{h}}_t}$ and backward $\overleftarrow{\textbf{\textit{h}}_t}$ directions
\begin{align}
\overrightarrow{\textbf{\textit{h}}_t} &= \sigma( \textbf{W}_{\overrightarrow{h}}  [ x_t, \, \overrightarrow{h}_{t-1} ] + \textbf{b}_{\overrightarrow{h}} ), \\
\overleftarrow{\textbf{\textit{h}}_t} &= \sigma( 
\textbf{W}_{\overleftarrow{h}}  [ x_t, \, \overleftarrow{h}_{t-1} ] + \textbf{b}_{\overleftarrow{h}} ) .
\end{align}
Then, parameters of the opposing hidden states are stacked $\overline{\textbf{\textit{h}}}_t$, and fed into the output layer to provide state denoising
\begin{align}
\overline{\textbf{\textit{h}}}_t &= [\overrightarrow{\textbf{\textit{h}}_t}, \overleftarrow{\textbf{\textit{h}}_t}] \ \in \ \mathbb{R}^{3 \times 2m}, \\
\hat{ \boldsymbol{x} }_t &= \boldsymbol{\phi}( \overline{\textbf{\textit{h}}}_t ) = \textbf{W}_{ \hat{y} }  \overline{\textbf{\textit{h}}}_t + \textbf{b}_{ \hat{y} } \ \in \ \mathbb{R}^3 \label{eq:stacked} .
\end{align}

\subsection{Bi-directional one-layer GRU} 
The third model is also arranged in a bi-directional form, but utilizes a different gating mechanism called gated recurrent unit (GRU), which merges both input and forget gates into a single update gate \cite{cho2014properties}. Using only two-gates, it requires shorter training epochs, less computational efforts and is thus more robust to the vanishing gradient problem. \textcolor{turquoise}{Using a sigmoid function, the gates regulate the trade-off between previous hidden states and new input information, as zero output means closed gate, such that historical data cannot pass beyond, and only the new input is emphasized.}
The reset gate $\textbf{\textit{r}}_t$ controls the dominance of past states with respect to new input, and the update gate $\textbf{\textit{z}}_t$ specifies how much memory will pass to the next hidden state. Since both hidden states are symmetrically opposite, only the chronological direction is formulated to simplify notation
\begin{align}
\textbf{\textit{r}}_t &= \sigma( \textbf{W}_r  [ x_t, \, \overrightarrow{h}_{t-1} ] + \textbf{b}_r  ), \\
\textbf{\textit{z}}_t &= \sigma( \textbf{W}_z  [ x_t, \, \overrightarrow{h}_{t-1} ] + \textbf{b}_z  ),
\end{align}
where the chronological hidden state is defined as 
\begin{align}
\tilde{\textbf{\textit{h}}}_t &= \tanh ( \textbf{W}_{\overrightarrow{h}} [ x_t, \ \textbf{\textit{r}}_t \odot \overrightarrow{h}_{t-1}] + \textbf{b}_{\overrightarrow{h}} ), \\
\overrightarrow{\textbf{\textit{h}}}_t &= (1- \textbf{\textit{z}}_t) \odot \overrightarrow{\textit{h}}_{t-1} + \textbf{\textit{z}}_t \odot \tilde{\textbf{\textit{h}}}_t .
\end{align}
Similarly to (\ref{eq:stacked}), the opposite cells are horizontally stacked and approximated by the external output layer as
\begin{align}
\overline{\textbf{\textit{h}}}_t &= [\overrightarrow{\textbf{\textit{h}}_t}, \overleftarrow{\textbf{\textit{h}}_t}] \ \in \ \mathbb{R}^{3 \times 2m} , \\
\hat{ \boldsymbol{x} }_t &= \boldsymbol{\phi}( \overline{\textbf{\textit{h}}}_t ) = \textbf{W}_{ \hat{y} }  \overline{\textbf{\textit{h}}}_t + \textbf{b}_{ \hat{y}} \ \in \ \mathbb{R}^3 .
\end{align}

\subsection{k-nearest neighbors algorithm (kNN)}
Our last proposed model does not assume priors on the underlying distribution, but is only determined by the number of $k$ nearest neighbors. Let the dataset be a set of $n$ pairs 
\begin{align} \label{eq:dpoints}
\mathcal{D} = \{( \boldsymbol{x}_i, \boldsymbol{x}_{_{GT},i} ) \}_{i=1}^n ,
\end{align}

where denoised outputs are determined by the mean GT value of $k$ closest (noisy) samples to the new data point
\begin{align} 
\hat{\boldsymbol{x}}(k) = \frac{1}{k} \sum_{ \boldsymbol{x}_i \in \boldsymbol{N}(k) }  \boldsymbol{x}_{_{GT},i}.
\end{align}

Given r-dimensional data points $\boldsymbol{x}_1$ and $\boldsymbol{x}_2$, neighborhood $\boldsymbol{N}$ will be consisted of $k$ closest neighbors, as proximity is calculated using an Euclidean distance function
\begin{align}
d( \boldsymbol{x}_1, \boldsymbol{x}_2 ) = \sqrt{ \sum_{ j=1 }^{r}  ( x_{1,j} - x_{2,j} )^2 } .
\end{align}

In the absence of pre-defined statistical methods to find optimality, an heuristic search brought up the following optimal $k$
 \begin{align} \label{eq:knn}
k^* = \arg \min_k \| \hat{\boldsymbol{X}}(k) - \boldsymbol{X}_{GT} \|^2 \approx \sqrt{n/5},
\end{align}
where $\hat{\boldsymbol{X}}$ is the kNN estimates and $\boldsymbol{X}_{GT}$ is their corresponding GT samples, both taken from the test-set, which is one fifth of the dataset ($n=90,000$).\\

\subsection{Complexity Analysis}
\textcolor{turquoise}{To estimate the feasibility of the proposed algorithms for real-time applications, Table 3. discusses the computational resources required, from theoretical and empirical points of view \cite{cormen2022introduction}. Using big O notation, time complexity provides an upper-bound estimate on the growth rate of the running time. As indicated in the third column, the relatively small model sizes, are due to an extensive heuristic optimization, enabling agility and usability for real-time applications. Finally, an estimated number of floating point operations (FLOPs) for a single forward pass, is given, referring to CPU-only inference.}

\begin{table}[ht]
\centering
\caption{Computational aspects of algorithms}
\renewcommand{\arraystretch}{1.2}
\begin{tabular}{c|c|c|c|}
Model & \shortstack{Time\\complexity} & \shortstack{Total\\params.} & \shortstack{Est.\\FLOPs} \\ \hline
MA  & $\mathcal{O}(W)$=$\mathcal{O}(1)$ & 1 & $< 1\text{\sc{e}}3$ \\ 
SG  & $\mathcal{O}( n d W^2 p )$ & 221 & $\approx 2\text{\sc{e}}4$ \\ 
DWT & $\mathcal{O}(n \log n \, d f)$ & 16 & $\approx 1\text{\sc{e}}4$ \\ \hline
Bi-RNN & $\vert$ & 9493 & $\approx 1\text{\sc{e}}5$ \\ 
LSTM & $\mathcal{O}(n d h)$ & 7323 & $\approx 8\text{\sc{e}}4$ \\ 
Bi-GRU & $\vert$ & 4653 & $\approx 5\text{\sc{e}}4$ \\ \hline
kNN & $\mathcal{O}(n^2 d )$ & $k^*$ & $\approx 3.5\text{\sc{e}}5$ \\  \hline
\end{tabular}
\label{t:complexity}
\end{table}
\textcolor{turquoise}{While the overall complexity is governed by the input size, $n$, other parameters, e.g. window size, $W$, polynomial order, $p$, filter length, $f$, hidden state size, $h$, and feature dimensionality, $d$, are constant in time, thus have a secondary impact on the scalability. As shown, all models perform fairly, i.e. $\leq \mathcal{O}(n \log n)$, except the kNN model. However using \eqref{eq:knn}, a reasonable run-time is guaranteed. The weak correlation between the number of parameters and the FLOPs is explained by unparameterized convolutions and matrix multiplications. As a rough estimate, modern multicore smartphones, capable of performing tens of billions of floating point operations per second (GFLOPS), would exhibit latency of maximum 1 ms.}

\section{Datasets and Error Measures} \label{sec:DnM}
To evaluate our methodology, models from both signal processing-based and learning-based approaches are evaluated and compared over simulated and experimental datasets, using unique error measures for the evaluation process. 



\subsection{Simulated Dataset}
In stationary conditions, the gravity vector is projected into the specific force measurements of the accelerometers. This projection can be represented using Euler angles. 
In our simulated setup, Euler angles are drawn within a bounded domain, and using small intervals, a rich combination of data points (orientations) is guaranteed. Using the transformation matrix (\ref{eq:T_b_n}), Euler angles dictate how gravity is projected on the sensor axes (\ref{eq:gravity}). Without loss of generality, the yaw angle has no influence on the gravity projection, thus it remains constant ($\psi_{sim.} = 0$), and we address only the roll and pitch angles in the range of 
\begin{align}
-15 ^{\circ} \, \leq \, & \phi_{sim.} \leq +15 ^{\circ}, \\
-15 ^{\circ} \, \leq \, & \theta_{sim.} \leq +15 ^{\circ} .
\end{align}

\begin{figure}[h]
\begin{center}
\includegraphics[width=0.6\textwidth]{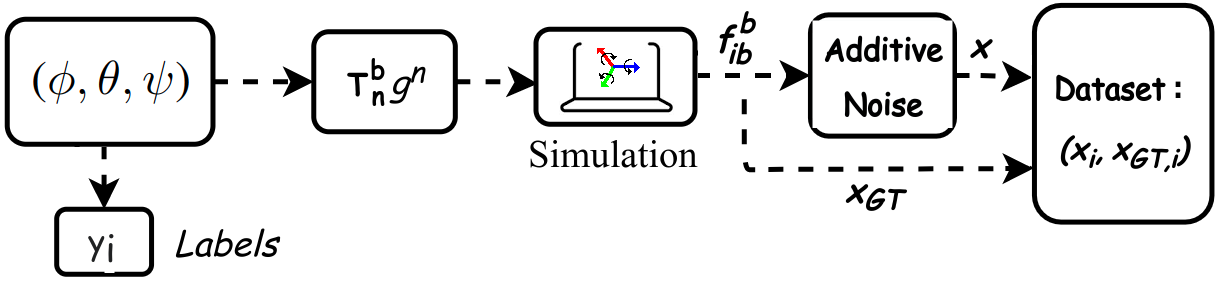}
\caption{Simulated dataset generation procedure}
\label{f:exp_sim}
\end{center}
\end{figure}

Both angles are divided into intervals of $0.1$ deg, such that $\frac{15-(-15)}{0.1}=300$ increments are obtained. Then, the angles are combined together, and the overall number of simulated instances becomes 300$^2$=90,000 combinations.
Notice, that in this study we limit the angles range to $\pm15 ^{\circ}$ to enable model training in fair times on our hardware (Intel i5-9600K CPU @ 3.70 GHz and NVIDIA GTX2080 GPU). Yet, the proposed methods can be applied in any angle range, influencing only the training time, and once trained, inference is applicable in real-time. Next, synthetic noise and error terms are added to each accelerometer channel using the following error terms:
\begin{enumerate}
   \item Velocity random walk (VRW) - accumulative error due to white noise in the measurement.
   \item Bias offset (BO) - a constant offset caused by misprojection of the true accelerations.
   \item Bias instability (BI) - stationary process which act as a low-order Gauss-Markov process.
\end{enumerate}
\textcolor{turquoiseee}{Although the first error is systematic, and the second is stochastic, both their characteristics are time invariant. In contrast, the random walk error is non-stationary, hardly modelable, thus threatening to violate the Gaussianity assumption. However, in short time periods below two minutes, higher order noise terms are assumed to remain dormant \cite{Groves2013, marinov2014allan}, such that the dynamic bias (BI) is assumed to be negligible compared to the static one (BO).} \\
Fig.~\ref{f:exp_sim} illustrates the simulation, showing that the constructed dataset contains both simulated GT and measured accelerometer readings. Table~\ref{t:noise} specifies the synthetic additive noise values (Simulation column).

\subsection{Field Experiment Dataset}
\begin{figure}[h]
\begin{center}
\includegraphics[width=0.45\textwidth]{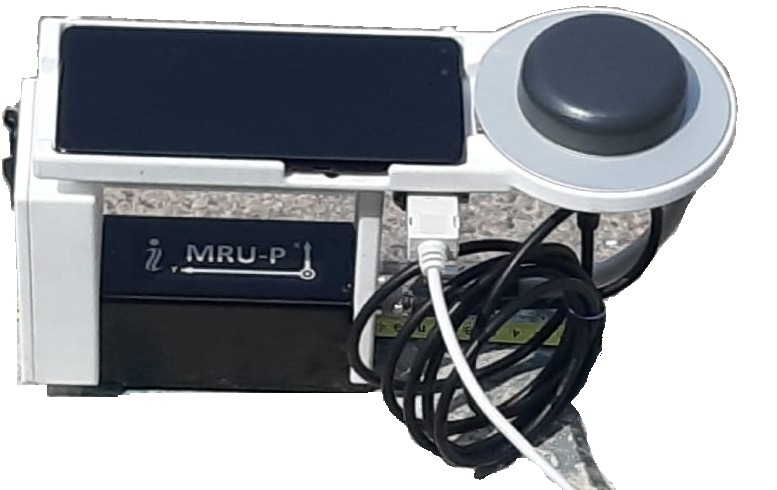}
\caption{A unique device aligning between a Huawei P40 smartphone and an Inertial Lab MRU-P unit.}
\label{f:Setup}
\end{center}
\end{figure}

The experimental data of the stationary conditions are based on a dataset published in \cite{shurin2022autonomous}, and the code execution is reproducible on the project page\footnote{For reproducibility, both data and code are publicly available @ \underline{\url{https://github.com/ANSFL/MEMS-IMU-Denoising}}.}. There, a unique device, as shown in Fig.~\ref{f:Setup} was built to align between a Huawei P40 smartphone and an Inertial Lab MRU-P unit \cite{MRU-ref}.

In this setup, the smartphone accelerometer readings are used as unit under test while the MRU's accelerometer readings serve as GT measurements. Both accelerometers provided measurements at 100Hz sampling rate. Due to practical constraints, smaller amount of raw measurements were taken in much bigger intervals of $\approx 3.0 ^{\circ}$ deg
\begin{align}
-15 ^{\circ} \, \leq \, & \phi_{exp.}   \leq +15 ^{\circ} ,
\\
-15 ^{\circ} \, \leq \, & \theta_{exp.}   \leq +15 ^{\circ} .
\end{align}
\textcolor{turquoise}{The experimental dataset consists of $10^2=100$ recordings in different orientations, with a maximum duration of two minutes. Than, each raw measurement is divided into non-overlapping sub-samples, depending on the chosen window size.}
Unlike the simulated dataset, yaw angles here are non-zeros ($\psi_{exp.} \neq 0$), as they were taken from a wide variety of platform orientations. 
Table~\ref{t:noise} specifies the sensors noise specs, as stated by manufacturers (Experiment column).
\begin{table}[ht]
\centering
\caption{Error sources used in the simulation and experiments}
\renewcommand{\arraystretch}{1.2}
\begin{tabular}{c|c|c|c|c|c|}
\multicolumn{2}{c}{} & \multicolumn{2}{c}{\textbf{Simulation}} & \multicolumn{2}{c}{\textbf{Experiment}} \\
Error & Units & Noisy & GT & P40  & MRU-P (GT) \\ \hline
VRW & [m/s/$\sqrt{s}$]  & 0.005 & 0.00001 & 0.003 & 0.00025 \\ \hline
BI & [m/s$^2$]          & 0.001 & 0.00001 & 0.001 & 0.00005 \\ \hline
BO & [m/s$^2$]          & 0.05 & 0.00001 & 0.067  & 0.0001 \\ \hline 
\end{tabular}
\label{t:noise}
\end{table}

Lastly, to increase variability of the experimental dataset and avoid model overfitting, each sample underwent two consecutive transformations suitable for time-series \cite{um2017data, ohashi2017augmenting, steven2018feature}: i) Angular augmentation - applying angular transformations over a given orientation, to obtain more different samples and densify the sparse distribution. ii) Noise augmentation - applying stochastic error in form of additive white Gaussian noise. This way, the augmented dataset also contains a total of 90,000 samples, improving generalizability and noise robustness of the learning-based models.

\subsection{Loss function}
Unlike the kNN optimization, deep learning architectures use a loss function to assess deviations between their denoised predictions and the actual ground truth. The axial error of a single sample containing $H$ time steps is given by 
\begin{align}
\boldsymbol{e}_{j} = \hat{\boldsymbol{x}}_{j} - \boldsymbol{x}_{j,_{GT}} \ \in \ \mathbb{R}^{H} ,
\label{eq:residual}
\end{align}
where $j$ denotes an axis index as each sensor measures an x-y-z triad. Then, during training, performance is assessed using mean squared error (MSE), such that loss surface is continuously differentiable, and errors can be minimized progressively as a function of the model weights 
\begin{align}
\text{MSE} = \frac{1}{H \times 3} \sum_{j \in \{\text{x,y,z}\}} \sum_{i=1}^H ( e_{i,j} )^2
\label{eq:mse}.
\end{align}

\subsection{Performance Metrics}
During the testing phase, several evaluation functions are used to provide relevant measures of the models performances from the perspective of signal reconstruction. 
\begin{itemize}
\item Root mean squared error (RMSE) - is defined as the square root of the MSE (\ref{eq:mse}), given by
\begin{equation} \label{eq:rmse}
\text{RMSE} = \sqrt{\text{MSE}}.
\end{equation}
\item Mean absolute error (MAE) - returns an average magnitude of the residuals (\ref{eq:residual}), calculated over $n$ time steps, disregarding their direction, given by
\begin{equation} \label{eq:mae}
\text{MAE} = \frac{1}{n} \sum_{i=1}^n |e_i|.
\end{equation}
\item Peak signal-to-noise ratio (PSNR) - returns the ratio between maximum signal value ($\text{MAX}$) and noise levels. However here, PSNR expresses ratio between two amplitudes (root-power quantity), thus a measure of order of magnitude is equal to 20 [dB] difference, given by
\begin{equation} \label{eq:psnr}
\text{PSNR} = 20 \log_{10} \left( \frac{ \text{MAX}}{\text{RMSE}} \right).
\end{equation}
\item Relative absolute error (RAE) - returns a relative measure of the discrepancy between the residuals and their corresponding ground truth mean value $\mu_{_{GT}}$, given by
\begin{equation} \label{eq:rae}
\text{RAE} = \frac{ \sum_{i=1}^n | e_i | }{ \sum_{i=1}^n | x_{_{GT},i} - \mu_{_{GT}} | } .
\end{equation}
\end{itemize}

\section{Analysis and Results}\label{sec:AandR}
This section presents a comparative performance analysis between all models introduced above: signal processing-based (\ref{subsec:sp}) and our proposed learning-based (\ref{sec:PA}). All measurements are assessed first in terms of pure signal reconstruction, then by comparing attitudes from the denoised outputs, as obtained by the stationary coarse alignment procedure. \textcolor{turquoiseee}{Note that the visualization windows are deliberately split into two time scales; macro-level, to emphasize the stationarity of the problem, and micro-level, to capture local nuances between the competing models, in a single time step.}
\begin{figure}[h]
\begin{center}
\includegraphics[width=0.675\textwidth]{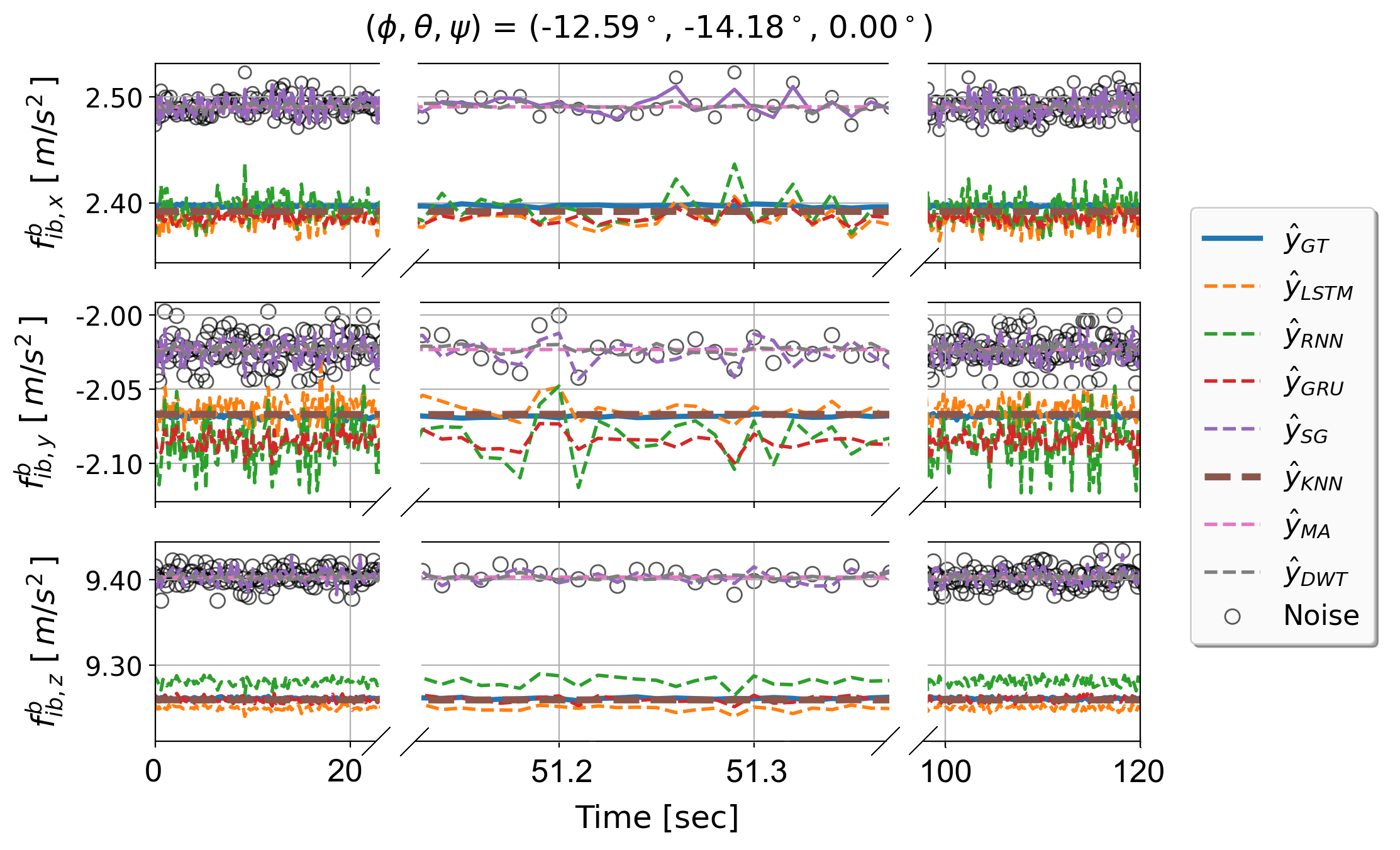} 
\caption{Reconstruction comparison: Simulation}
\label{f:sim_recon}
\end{center}
\end{figure}

\subsection{Simulation Assessment}
After the simulated dataset is generated, models are trained on the training set before comparing performances. Then, the simulated test-set utilizes to examine the models capability of reducing noise sources, in terms of error minimization. Fig.~\ref{f:sim_recon} visualizes model inference on a single noisy sample (bubbles), where GT measurements are denoted by solid blue line and models estimates $\hat{\boldsymbol{x}}$ are marked with dashed lines. 
Since this dataset is simulated, different tradeoffs between deterministic and stochastic noise sources were tested to examine models robustness. In the absence of a calibration process or knowledge about inherent bias, SP-based performances remained bounded by the sensors offset errors. In contrast, the learning-based estimators are optimized during training phase, capable of compensating wide range of error sources. 
\\
Table \ref{t:sim_comparison} summarizes the models denoising performances across the simulated training set, providing dissimilarity measures between reconstructed signals and the GT observations. 
\begin{table}[h] 
\caption{Reconstruction comparison: Simulation}
\centering
\renewcommand{\arraystretch}{1.15}
\begin{tabular}{c|c|c|c|c|}
Model & RMSE [m/s$^2$]& MAE [m/s$^2$]& PSNR [dB]& RAE [$\%$] \\ \hline
kNN 	& 0.02481 & 0.01912 & 52.10719 & 0.44484  \\ \hline 
GRU 	& 0.02635 & 0.01950 & 51.39078 & 0.46668  \\ 
LSTM 	& 0.02701 & 0.01999 & 50.10601 & 0.47835  \\ 
RNN 	& 0.02898 & 0.02145 & 46.25170 & 0.51335  \\ 
DWT 	& 0.06278 & 0.04623 & 43.81552 & 1.11545  \\ 
SG 	    & 0.06298 & 0.04678 & 43.62187 & 1.11605  \\ 
MA 	    & 0.06344 & 0.04872 & 43.42730 & 1.12842  \\ \hline  
Noisy 	& 0.06614 & 0.04913 & 41.73747 & 1.18383  \\ \hline 
\end{tabular}
\label{t:sim_comparison}
\end{table}

Models are ranked according to RMSE results, where learning-based denoisers have a clear advantage over the SP-based. Since kNN is leading the table, a suppression ratio $\gamma$ is defined, to quantify its ability to minimize the reconstruction error, with respect to raw noisy measurements 
\begin{align}
\gamma_{\text{sim.}} = \frac{\text{RAE}_{\text{kNN}}}{\text{RAE}_{\text{Noisy}}} \approx 37.5 \% .
\label{eq:sup_sim}
\end{align}

As seen, the relative absolute error (RAE) of the kNN estimates, cuts dissimilarity error in more than half when compared to noisy measurements. Generalization error is minimized the closer the output reaches the GT, but increases proportionally with respect to sudden noise bursts. 
Next, the denoised signals in Fig.~\ref{f:sim_recon} are computed through the analytic SCA equations (\ref{eq:roll}, \ref{eq:pitch}), to examine whether the reconstructed signals improve the roll and pitch accuracy. 
\begin{figure}[h]
\begin{center}
\includegraphics[width=0.65\textwidth]{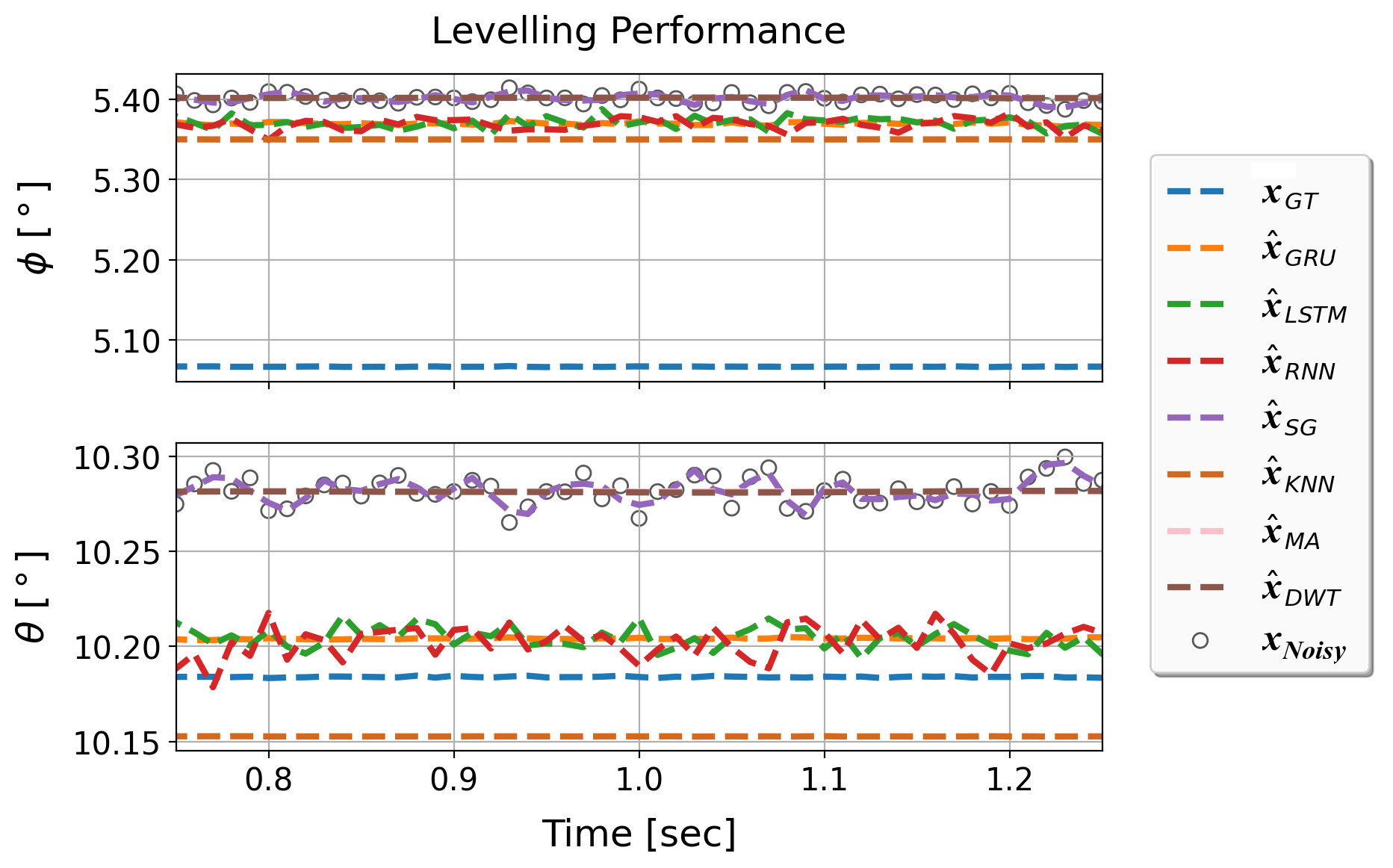}
\caption{SCA comparison: Simulation}
\label{f:sim_sca}
\end{center}
\end{figure}
Fig.~\ref{f:sim_sca} demonstrates how the computed angles obtained from models outputs, improve the SCA procedure, as computed roll and pitch angles lay closer to the ground-truth. Since the kNN is leading the accuracy, the next definition is used to assess the angular error between computed angle and GT angle
\begin{align}
\varepsilon = \alpha - \alpha_{GT} \ \in \ \mathbb{R} .
\end{align}

\begin{figure}[!h]
\begin{center}
\includegraphics[width=0.7\textwidth]{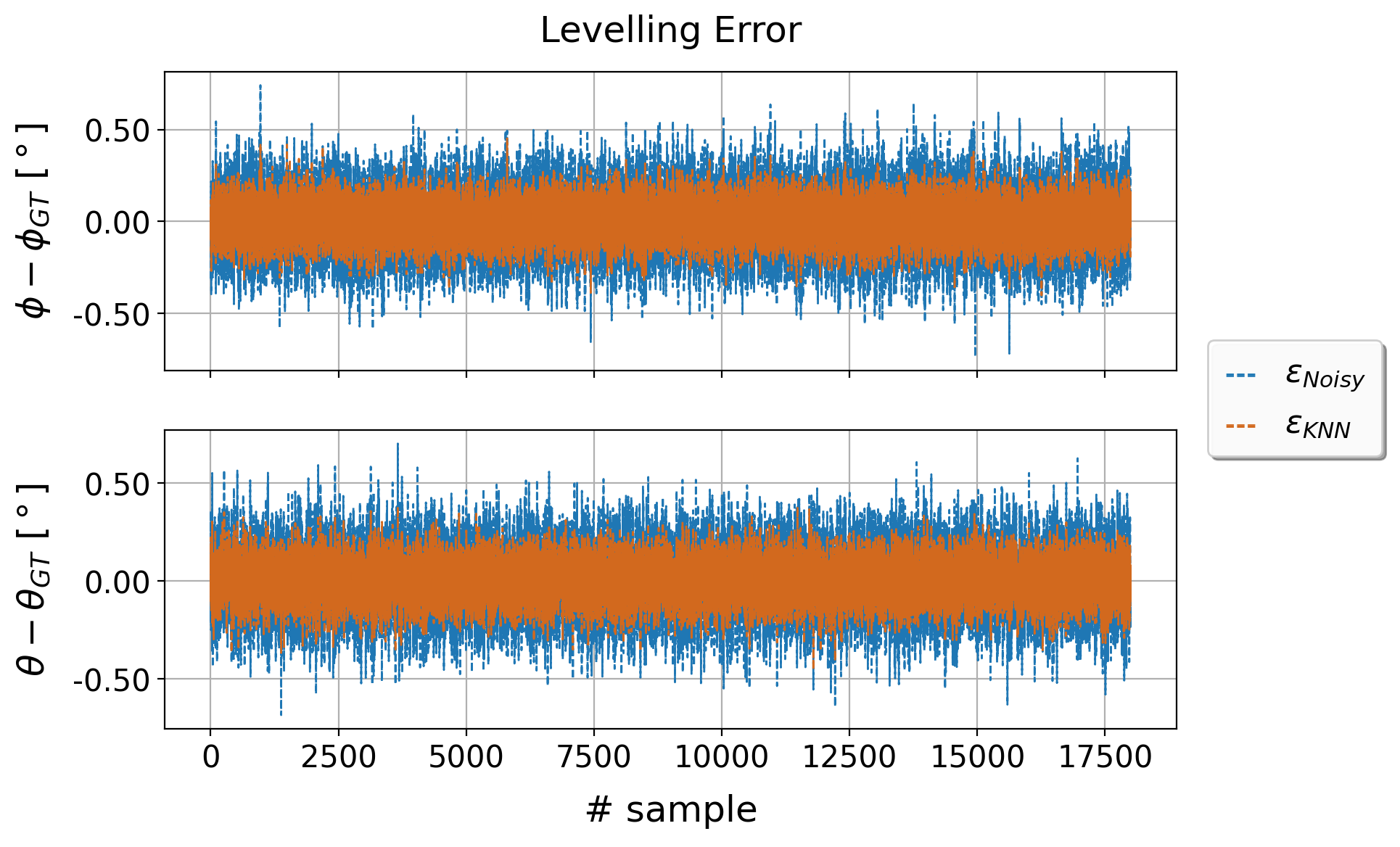}
\caption{SCA results of kNN: Simulation}
\label{f:sim_testset}
\end{center}
\end{figure}

Fig.~\ref{f:sim_testset} demonstrates the angular errors with respect to GT, as kNN estimated angles (brown) are compared with noisy raw measurements (blue), over the entire simulated test-set.

To understand the meaning of these results, Table \ref{t:sim_error} utilizes RMSE as a performance indicator, to summarize the kNN contribution to lowering the overall angular error. 
\begin{table}[h] 
\caption{Overall improvement rate: Simulation}
\centering
\renewcommand{\arraystretch}{1.15}
\begin{tabular}{c|c|c|c|}
  & $\varepsilon_{\text{Noisy}} \ [^\circ]$ & $\varepsilon_{\text{kNN}} \ [^\circ]$ & ratio [$\%$] \\ \hline
RMSE($\phi$) & 0.18452 & 0.14895 & 80.695  \\ \hline 
RMSE($\theta$) & 0.18147 & 0.14167 & 78.068  \\ \hline
\end{tabular}
\label{t:sim_error}
\end{table}

To conclude, synthetic noises and biases were generated to simulate contaminated measurements and to examine the models capabilities to remove them. Empirical experiments showed that each architecture exhibited different bias-variance tradeoffs. Some estimators performed better using shallow architectures such that variance error (overfitting) was reduced. Others improved when estimated parameters were more noise-sensitive, thus avoiding underfitting due to sensors biases. Over the entire test-set, it was shown that learning-based estimates reduce reconstruction error by up to 37.5$\%$, followed by 20$\%$ accuracy improvement of the SCA procedure. 
%

\subsection{Experimental Assessment} \label{subsec:exp}
\begin{figure}[!h]
\begin{center}
\includegraphics[width=0.675\textwidth]{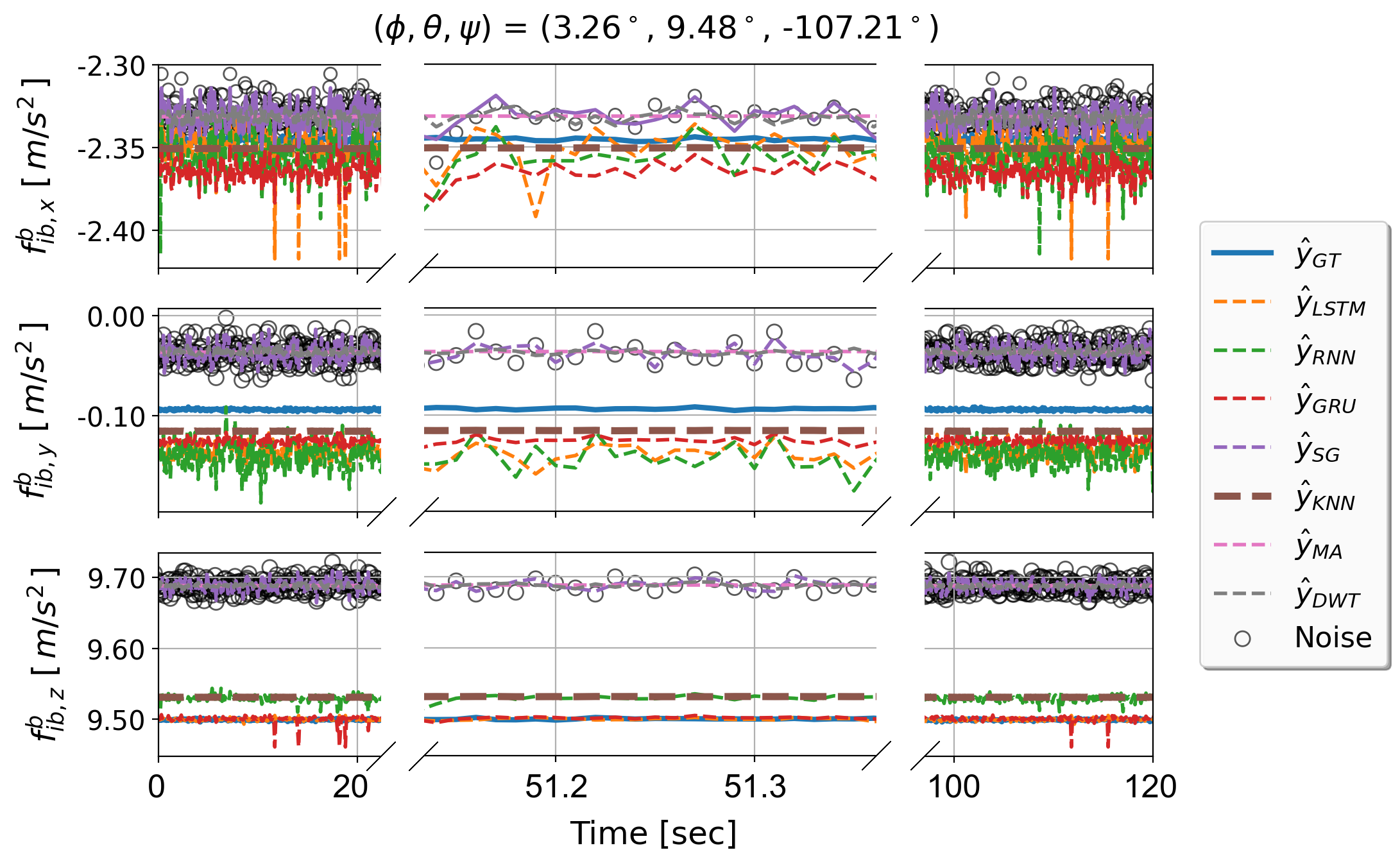} 
\caption{Reconstruction comparison: Experiment}
\label{f:real_recon}
\end{center}
\end{figure}

Unlike the simulated dataset, error sources in the experimental setup are sensor-specific, and tradeoff between them cannot be modified. 
Here, noisy samples are given by a consumer-grade smartphone sensor, whose readings are submerged in lower levels of stochastic noise but higher bias levels, as smartphones can be prone to mechanical shocks that impair the orthogonality of the sensor axes. In contrast, the GT references are taken from a high-end sensor, whose accurate measurements are characterized by significantly lower noise levels \cite{shurin2022autonomous}. In other words, the generalization task here is to find an approximation function that mimics best the GT sensor. Similarly to before, Fig.~\ref{f:real_recon} compares the denoising capabilities of each model, as the optimization task is to minimize dissimilarity between the model estimates and GT.

When compared to the simulated dataset, noise (bubbles) exhibits more bias but less stochastic noise, as fluctuations are smaller. Here as well, the learning-based algorithms lay significantly closer to GT measurements, outperforming the SP-based models which lay in the vicinity of the noisy measurements. It is not unusual however, as conventional filters may excel at spatial filtering, but in the absence of a calibration process, constant biases remain unfiltered. 
Table \ref{t:real_comparison} summarizes the models reconstruction errors when computed across the entire experimental test-set.
\begin{table}[h] 
\caption{Reconstruction comparison: Experiment}
\centering
\renewcommand{\arraystretch}{1.15}
\begin{tabular}{c|c|c|c|c|}
Model & RMSE [m/s$^2$]& MAE [m/s$^2$]& PSNR [dB]& RAE [$\%$] \\ \hline
kNN 	& 0.00969 & 0.00658 & 60.09720 & 0.171414  \\ \hline 
GRU 	& 0.01373 & 0.01016 & 57.07227 & 0.242834  \\
LSTM 	& 0.01414 & 0.01082 & 56.82121 & 0.249957  \\
RNN 	& 0.01505 & 0.01231 & 56.27788 & 0.266084  \\
SG 	    & 0.10317 & 0.09124 & 39.81290 & 1.727121  \\ 
DWT 	& 0.10326 & 0.09126 & 39.55169 & 1.823734  \\
MA 	    & 0.10588 & 0.09396 & 39.52071 & 1.825294  \\ \hline 
Noisy	& 0.10843 & 0.09427 & 39.53768 & 1.828244  \\ \hline 
\end{tabular}
\label{t:real_comparison}
\end{table}

\begin{figure}[h]
\begin{center}
\includegraphics[width=0.65\textwidth]{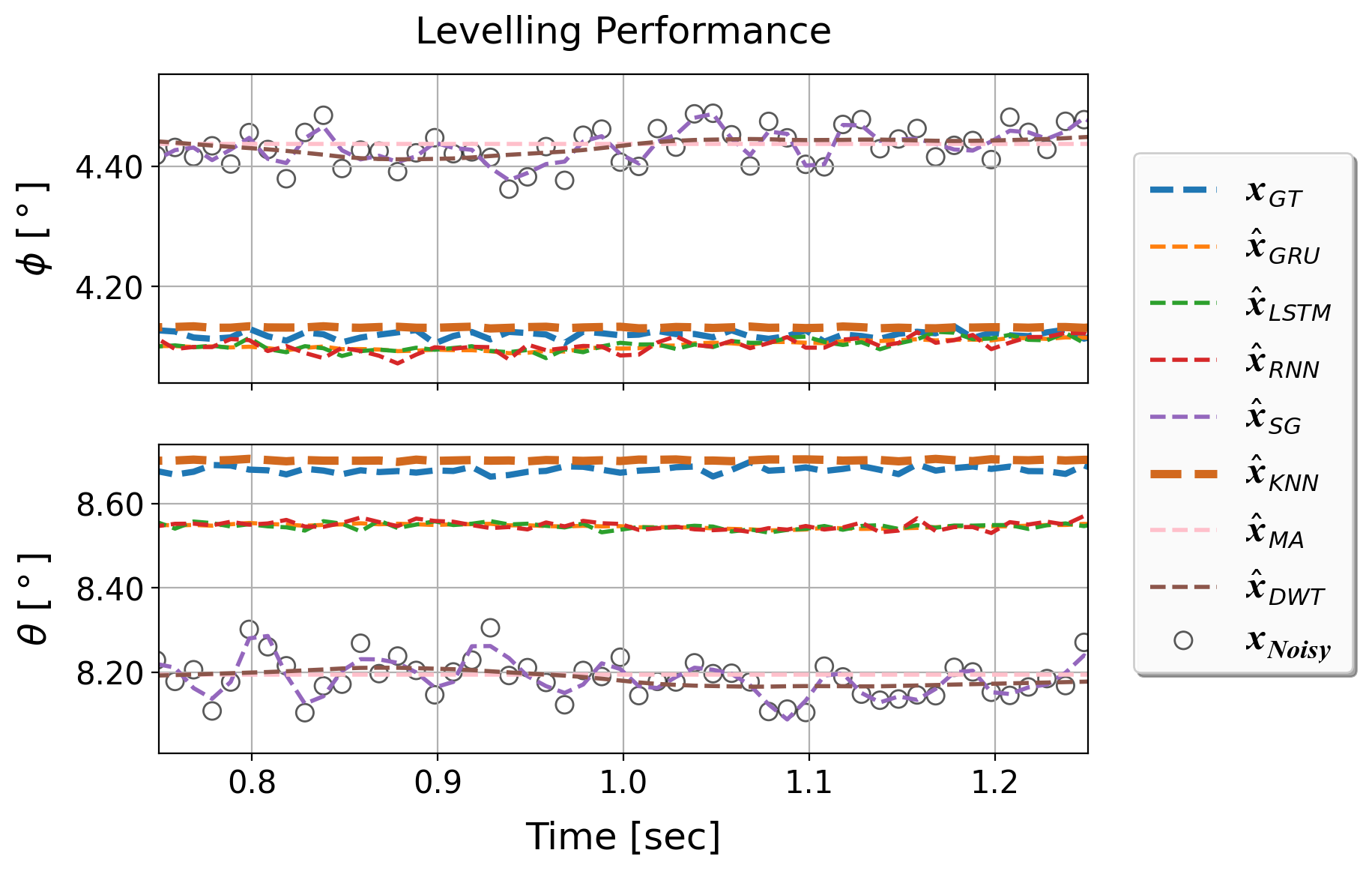}
\caption{SCA comparison: Experiment}
\label{f:real_levelling}
\end{center}
\end{figure}

\begin{figure}[!h]
\begin{center}
\includegraphics[width=0.67\textwidth]{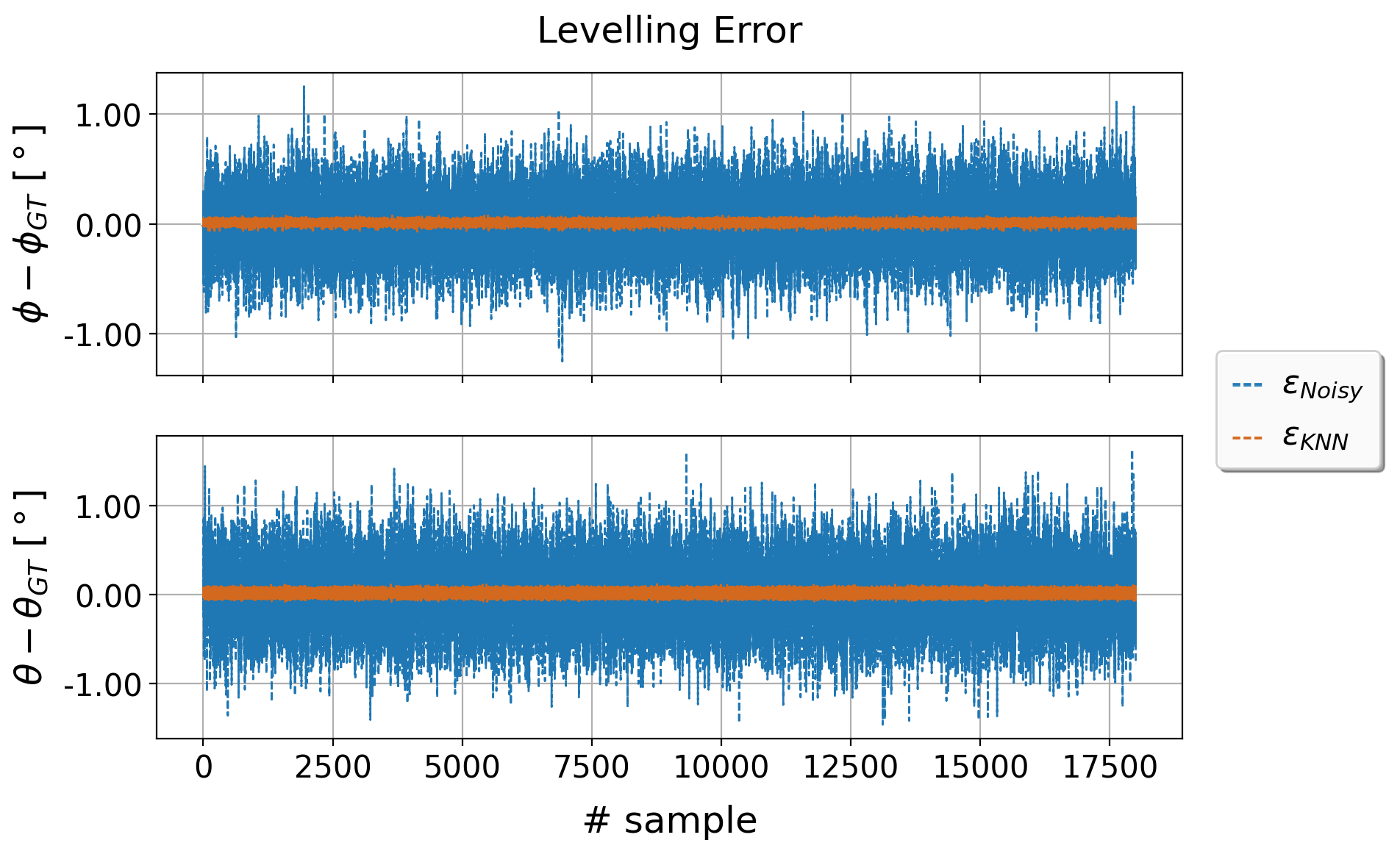}
\caption{SCA results of kNN: Experiment}
\label{f:overall_SCA}
\end{center}
\end{figure}

Once again, the kNN model demonstrates superiority among all models, as its relative absolute error reduces error by an order of magnitude, as given by the suppression ratio 
\begin{align}
\gamma_{exp.} = \frac{\text{RAE}_{\text{kNN}}}{\text{RAE}_{\text{Noisy}}} \approx 9.38 \%
\label{eq:exp}
\end{align}
The strong differences between $\gamma_{sim.}$ and $\gamma_{exp.}$ can be explained by unknown error sources that are present in real-world devices, but underestimated or mismodeled during the simulation phase, thus beneficial to the experimental results.
Next, the reconstructed signals shown in Fig.~\ref{f:real_recon}, are used to estimate the roll and pitch angles of the SCA procedure, as shown in Fig.~\ref{f:real_levelling}.
The learning-based denoisers exhibit better performances than the SP-based, as their outputs manage to minimize the distance error with respect to the GT angles. Focusing again on the best estimator, Fig.~\ref{f:overall_SCA} shows angular errors of computed roll and pitch angles, comparing noisy measurements (blue) with kNN estimates (brown), across the experimental test-set.
Despite greater noise amplitude of the instruments, the kNN model manages to reduce reconstruction errors drastically.
Table \ref{t:real_error} shows the overall improvement rate of the computed angles, using RMSE performance indicators to compare the kNN contribution with respect to noisy sensor readings. 
\begin{table}[h] 
\caption{Overall improvement rate: Experiment}
\centering
\renewcommand{\arraystretch}{1.05}
\begin{tabular}{c|c|c|c|}
  & $\varepsilon_{\text{Noisy}} \ [^\circ]$ & $\varepsilon_{\text{kNN}} \ [^\circ]$ & ratio [$\%$] \\ \hline
RMSE($\phi$) & 0.30042 & 0.03398 & 11.311  \\ \hline 
RMSE($\theta$) & 0.50251 & 0.06676 & 13.285  \\ \hline
\end{tabular}
\label{t:real_error}
\end{table}

Similarly to the reconstruction results (\ref{eq:exp}), the angular errors are also reduced by one order of magnitude, confirming the contribution of learning-based denoising techniques.
To conclude the experimental section, accelerometer readings from a commercial-grade smartphone sensor were used as noisy measurements, whereas accurate readings from an aligned high-end sensor were used as a GT reference. The denoising capabilities of different models were investigated, followed by examining their contribution to the SCA procedure. \\
Similarly to the simulated scenario, our proposed kNN algorithm happened to outperform all other models, managing to reduce the angular errors by one order of magnitude. We explain its superiority by the simplistic non-parametric approach, which relies on semantic similarity between independent variables, here specific force measurements, with respect to $k$ nearest data points. 

\section{Limitations of the study} \label{sec:limit}
\textcolor{turquoiseee}{
While the study at hand exhibited the unprecedented capabilities of the proposed learning-based models, it is important to acknowledge several limitations that naturally arise:
\begin{enumerate}
\item Generalizability: The study examined the ability of learning-approaches to cope with noisy and biased data. As such, models rely heavily on the labeling quality, and quantity, during the training phase. Previously unseen data during the inference stage, in form of unfamiliar dynamics or noise pattern, may degrade their superiority over the SP-based methods.
\item Usability: The authors admit that some conventional methods can be found significantly simpler to operate, offering training-free implementation, thus providing predictions right away. 
\item Interpretability: In contrast to SP-based methods, and despite their approximation capabilities, deep learning models often act as black boxes, as their underlying mapping function remains latent, hardly explainable, with an unclear decision-making process.
\end{enumerate} }

\section{Conclusions}\label{sec:CN}
In this work we addressed the challenging task of stationary accelerometer denoising using data-driven methods, over a simulated and an experimental datasets. By implementing and adjusting a wide range of denoising algorithms, we assessed their capabilities in terms of pure noise suppression, followed by examining the SCA improvement rates. The results show a clear advantage of the learning-based methods over the conventional signal processing algorithms, presumably due to their ability to compensate a wide range of error sources. \\
Yet, improvement rates seemed to vary significantly between both experiments. We explain this by the simulated scenario which enabled us the freedom of choice to set the intensity of the noisy samples, and vice verse, to the determine GT errors to be almost noiseless. While in the simulated assessment, characteristic errors dropped to half, in the experimental assessment they were reduced by more than one order of magnitude. 
\\
In light of the promising results obtained in stationary conditions, future research regarding inertial denoising under dynamic conditions is required. The question in focus should first determine whether learning-based approaches are even suitable for this type of tasks. If so, will they manage to generalize well given complex dynamics, and still get the upper hand over conventional approaches.

\bibliographystyle{elsarticle-num} 
\bibliography{cas-refs}

\begin{thebibliography}{10}
\expandafter\ifx\csname url\endcsname\relax
  \def\url#1{\texttt{#1}}\fi
\expandafter\ifx\csname urlprefix\endcsname\relax\def\urlprefix{URL }\fi
\expandafter\ifx\csname href\endcsname\relax
  \def\href#1#2{#2} \def\path#1{#1}\fi

\bibitem{Jekeli2000}
C.~Jekeli, Inertial navigation systems with geodetic applications, Walter de
  Gruyter Berlin, Germany, 2000.

\bibitem{Groves2013}
P.~D. Groves, Principles of {GNSS}, Inertial and Multisensor Integrated
  Navigation Systems, Artech House, 2013.

\bibitem{Titterton2004}
D.~Titterton, J.~L. Weston, Strapdown Inertial Navigation Technology, American
  Institute of Aeronautics and Astronautics and the Institution of Electrical
  Engineers, 2004.

\bibitem{el2007analysis}
N.~El-Sheimy, H.~Hou, X.~Niu, Analysis and modeling of inertial sensors using
  allan variance, IEEE Transactions on instrumentation and measurement 57~(1)
  (2007) 140--149.

\bibitem{dadafshar2014accelerometer}
M.~Dadafshar, Accelerometer and gyroscopes sensors: operation, sensing, and
  applications, Maxim Integrated [online] (2014).

\bibitem{gonzalez2018statistical}
R.~Gonzalez, C.~A. Catania, A statistical approach for optimal order adjustment
  of a moving average filter, in: 2018 IEEE/ION Position, Location and
  Navigation Symposium (PLANS), IEEE, 2018, pp. 1542--1546.

\bibitem{waegli2010noise}
A.~Waegli, J.~Skaloud, S.~Guerrier, M.~E. Par{\'e}s, I.~Colomina, Noise
  reduction and estimation in multiple micro-electro-mechanical inertial
  systems, Measurement Science and Technology 21~(6) (2010) 065201.

\bibitem{diao2013analysis}
Z.~Diao, H.~Quan, L.~Lan, Y.~Han, Analysis and compensation of {MEMS} gyroscope
  drift, in: 2013 Seventh International Conference on Sensing Technology
  (ICST), IEEE, 2013, pp. 592--596.

\bibitem{yong2015research}
S.~Yong, C.~Jiabin, S.~Chunlei, H.~Yongqiang, Research on the compensation in
  {MEMS} gyroscope random drift based on time-series analysis and kalman
  filtering, in: 2015 34th Chinese Control Conference (CCC), IEEE, 2015, pp.
  2078--2082.

\bibitem{tu2020arma}
Y.-H. Tu, C.-C. Peng, An {ARMA}-based digital twin for {MEMS} gyroscope drift
  dynamics modeling and real-time compensation, IEEE Sensors Journal 21~(3)
  (2020) 2712--2724.

\bibitem{abbasi2022memory}
J.~Abbasi, M.~Hashemi, A.~Alasty, A memory-based filter for long-term error
  de-noising of {MEMS}-gyros, IEEE Transactions on Instrumentation and
  Measurement 71 (2022) 1--8.

\bibitem{nassar2004modeling}
S.~Nassar, K.-P. Schwarz, A.~Noureldin, et~al., Modeling inertial sensor errors
  using autoregressive ({AR}) models, NAVIGATION, Journal of the Institute of
  Navigation 51~(4) (2004) 259--268.

\bibitem{nassar2005accurate}
S.~Nassar, Accurate insidgps positioning using {INS} data de-noising and
  auto-regressive modeling of inertial sensor errors, Geomatica 59~(3) (2005)
  283--294.

\bibitem{wang2018time}
X.~Wang, L.~Wang, Time-serial modeling and kalman filter of {MEMS} gyroscope
  random drift compensation, in: 2018 IEEE CSAA Guidance, Navigation and
  Control Conference (CGNCC), IEEE, 2018, pp. 1--5.

\bibitem{yuan2010research}
G.~Yuan, H.~Liang, K.~He, Y.~Xie, Research on signal de-noising technique for
  {MEMS} gyro, in: 2010 3rd International Symposium on Systems and Control in
  Aeronautics and Astronautics, IEEE, 2010, pp. 1288--1291.

\bibitem{gan2014emd}
Y.~Gan, L.~Sui, J.~Wu, B.~Wang, Q.~Zhang, G.~Xiao, An {EMD} threshold
  de-noising method for inertial sensors, Measurement 49 (2014) 34--41.

\bibitem{liu2019gyroscope}
C.~Liu, Z.~Yang, Z.~Shi, J.~Ma, J.~Cao, A gyroscope signal denoising method
  based on empirical mode decomposition and signal reconstruction, Sensors
  19~(23) (2019) 5064.

\bibitem{wang2021research}
X.~Wang, H.~Cao, Y.~Jiao, T.~Lou, G.~Ding, H.~Zhao, X.~Duan, Research on novel
  denoising method of variational mode decomposition in mems gyroscope,
  Measurement Science Review 21~(1) (2021) 19--24.

\bibitem{shen2016noise}
C.~Shen, J.~Li, X.~Zhang, Y.~Shi, J.~Tang, H.~Cao, J.~Liu, A noise reduction
  method for dual-mass micro-electromechanical gyroscopes based on sample
  entropy empirical mode decomposition and time-frequency peak filtering,
  Sensors 16~(6) (2016) 796.

\bibitem{guo2018hybrid}
X.~Guo, C.~Sun, P.~Wang, L.~Huang, Hybrid methods for {MEMS} gyro signal noise
  reduction with fast convergence rate and small steady-state error, Sensors
  and Actuators A: Physical 269 (2018) 145--159.

\bibitem{liu2020denoising}
Y.~Liu, G.~Chen, Z.~Wei, J.~Yang, D.~Xing, Denoising method of {MEMS} gyroscope
  based on interval empirical mode decomposition, Mathematical Problems in
  Engineering 2020 (2020).

\bibitem{li2013noise}
Q.~Li, X.~Chen, W.~Xu, Noise reduction of accelerometer signal with singular
  value decomposition and savitzky-golay filter, JOURNAL OF INFORMATION
  \&COMPUTATIONAL SCIENCE 10~(15) (2013) 4783--4793.

\bibitem{nirmal2016noise}
K.~Nirmal, A.~Sreejith, J.~Mathew, M.~Sarpotdar, A.~Suresh, A.~Prakash,
  M.~Safonova, J.~Murthy, Noise modeling and analysis of an {IMU}-based
  attitude sensor: improvement of performance by filtering and sensor fusion,
  in: Advances in Optical and Mechanical Technologies for Telescopes and
  Instrumentation II, Vol. 9912, International Society for Optics and
  Photonics, 2016, p. 99126W.

\bibitem{karaim2019low}
M.~Karaim, A.~Noureldin, T.~B. Karamat, Low-cost {IMU} data denoising using
  savitzky-golay filters, in: 2019 International Conference on Communications,
  Signal Processing, and their Applications (ICCSPA), IEEE, 2019, pp. 1--5.

\bibitem{he2019noise}
J.~He, C.~Sun, P.~Wang, Noise reduction for mems gyroscope signal: a novel
  method combining acmp with adaptive multiscale sg filter based on ama,
  Sensors 19~(20) (2019) 4382.

\bibitem{kang2010wavelet}
C.~W. Kang, C.~H. Kang, C.~G. Park, Wavelet denoising technique for improvement
  of the low cost {MEMS-GPS} integrated system, Wiley New York (2010).

\bibitem{kang2011improvement}
C.-H. Kang, S.-Y. Kim, C.-G. Park, Improvement of a low cost {MEMS}
  inertial-{GPS} integrated system using wavelet denoising techniques,
  International Journal of Aeronautical and Space Sciences 12~(4) (2011)
  371--378.

\bibitem{el2004wavelet}
N.~El-Sheimy, S.~Nassar, A.~Noureldin, Wavelet de-noising for {IMU} alignment,
  IEEE Aerospace and Electronic Systems Magazine 19~(10) (2004) 32--39.

\bibitem{liu2007mems}
F.~Liu, F.~Liu, W.~Wang, B.~Xu, {MEMS} gyro's output signal de-noising based on
  wavelet analysis, in: 2007 International Conference on Mechatronics and
  Automation, IEEE, 2007, pp. 1288--1293.

\bibitem{li2014improved}
Z.-p. Li, Q.-j. Fan, L.-m. Chang, X.-h. Yang, Improved wavelet threshold
  denoising method for mems gyroscope, in: 11th IEEE International Conference
  on Control \& Automation (ICCA), IEEE, 2014, pp. 530--534.

\bibitem{song2019mems}
J.~Song, Z.~Shi, B.~Du, L.~Han, H.~Wang, Z.~Wang, Mems gyroscope wavelet
  de-noising method based on redundancy and sparse representation,
  Microelectronic Engineering 217 (2019) 111112.

\bibitem{qu2009adaptive}
G.~Qu, F.~Zhao, G.~Liu, H.~Liu, Adaptive {MEMS} gyroscope denoising method
  based on the {\`a} trous wavelet transform, in: 2009 9th International
  Conference on Electronic Measurement \& Instruments, IEEE, 2009, pp. 2--787.

\bibitem{yuan2015improved}
J.~Yuan, Y.~Yuan, F.~Liu, Y.~Pang, J.~Lin, An improved noise reduction
  algorithm based on wavelet transformation for {MEMS} gyroscope, Frontiers of
  Optoelectronics 8~(4) (2015) 413--418.

\bibitem{el2018utilization}
A.~S. El-Wakeel, A.~Noureldin, H.~S. Hassanein, N.~Zorba, Utilization of
  wavelet packet sensor de-noising for accurate positioning in intelligent road
  services, in: 2018 14th International Wireless Communications \& Mobile
  Computing Conference (IWCMC), IEEE, 2018, pp. 1231--1236.

\bibitem{Ali_2021}
T.~A. Ali, A.~M. Hasan,
  \href{https://doi.org/10.1088/1757-899x/1094/1/012066}{A wavelet-{NARX} model
  for {SDINS}/{GPS} integration system}, {IOP} Conference Series: Materials
  Science and Engineering 1094~(1) (2021) 012066.
\newblock \href {https://doi.org/10.1088/1757-899x/1094/1/012066}
  {\path{doi:10.1088/1757-899x/1094/1/012066}}.
\newline\urlprefix\url{https://doi.org/10.1088/1757-899x/1094/1/012066}

\bibitem{khanafer2020applied}
M.~Khanafer, S.~Shirmohammadi, Applied {AI} in instrumentation and measurement:
  The deep learning revolution, IEEE Instrumentation \& Measurement Magazine
  23~(6) (2020) 10--17.

\bibitem{gonzalez2019time}
R.~Gonzalez, C.~A. Catania, Time-delayed multiple linear regression for
  de-noising {MEMS} inertial sensors, Computers \& Electrical Engineering 76
  (2019) 1--12.

\bibitem{brossard2020denoising}
M.~Brossard, S.~Bonnabel, A.~Barrau, Denoising {IMU} gyroscopes with deep
  learning for open-loop attitude estimation, IEEE Robotics and Automation
  Letters 5~(3) (2020) 4796--4803.

\bibitem{huang2022mems}
F.~Huang, Z.~Wang, L.~Xing, C.~Gao, A {MEMS IMU} gyroscope calibration method
  based on deep learning, IEEE Transactions on Instrumentation and Measurement
  71 (2022) 1--9.

\bibitem{jiang2018mems}
C.~Jiang, S.~Chen, Y.~Chen, B.~Zhang, Z.~Feng, H.~Zhou, Y.~Bo, A {MEMS IMU}
  de-noising method using long short term memory recurrent neural networks
  ({LSTM-RNN}), Sensors 18~(10) (2018) 3470.

\bibitem{jiang2018performance}
C.~Jiang, S.~Chen, Y.~Chen, Y.~Bo, L.~Han, J.~Guo, Z.~Feng, H.~Zhou,
  Performance analysis of a deep simple recurrent unit recurrent neural network
  ({SRU-RNN}) in {MEMS} gyroscope de-noising, Sensors 18~(12) (2018) 4471.

\bibitem{ruoyu2019modeling}
Z.~Ruoyu, G.~Shuang, C.~Xiaowen, Modeling of {MEMS} gyro drift based on wavelet
  threshold denoising and improved elman neural network, in: 2019 14th IEEE
  International Conference on Electronic Measurement \& Instruments (ICEMI),
  IEEE, 2019, pp. 1754--1761.

\bibitem{zhu2019mems}
Z.~Zhu, Y.~Bo, C.~Jiang, A mems gyroscope noise suppressing method using neural
  architecture search neural network, Mathematical Problems in Engineering 2019
  (2019).

\bibitem{zhu2021combined}
C.~Zhu, S.~Cai, Y.~Yang, W.~Xu, H.~Shen, H.~Chu, A combined method for {MEMS}
  gyroscope error compensation using a long short-term memory network and
  kalman filter in random vibration environments, Sensors 21~(4) (2021) 1181.

\bibitem{han2021hybrid}
S.~Han, Z.~Meng, X.~Zhang, Y.~Yan, Hybrid deep recurrent neural networks for
  noise reduction of mems-imu with static and dynamic conditions, Micromachines
  12~(2) (2021) 214.

\bibitem{jiang2019mixed}
C.~Jiang, Y.~Chen, S.~Chen, Y.~Bo, W.~Li, W.~Tian, J.~Guo, A mixed deep
  recurrent neural network for {MEMS} gyroscope noise suppressing, Electronics
  8~(2) (2019) 181.

\bibitem{schmidhuber2015deep}
J.~Schmidhuber, Deep learning in neural networks: An overview, Neural networks
  61 (2015) 85--117.

\bibitem{elman1990finding}
J.~L. Elman, Finding structure in time, Cognitive science 14~(2) (1990)
  179--211.

\bibitem{hochreiter1997long}
S.~Hochreiter, J.~Schmidhuber, Long short-term memory, Neural computation 9~(8)
  (1997) 1735--1780.

\bibitem{cho2014properties}
K.~Cho, B.~Van~Merri{\"e}nboer, D.~Bahdanau, Y.~Bengio, On the properties of
  neural machine translation: Encoder-decoder approaches, arXiv preprint
  arXiv:1409.1259 (2014).

\bibitem{cormen2022introduction}
T.~H. Cormen, C.~E. Leiserson, R.~L. Rivest, C.~Stein, Introduction to
  algorithms, MIT press, 2022.

\bibitem{marinov2014allan}
M.~Marinov, Z.~Petrov, Allan variance analysis on error characters of low-cost
  mems accelerometer mma8451q, in: International conference of scientific paper
  AFASES, 2014, pp. 22--24.

\bibitem{shurin2022autonomous}
A.~Shurin, A.~Saraev, M.~Yona, Y.~Gutnik, S.~Faber, A.~Etzion, I.~Klein, The
  autonomous platforms inertial dataset, IEEE Access 10 (2022) 10191--10201.

\bibitem{MRU-ref}
I.~Labs, {Motion Reference Unit},
  \url{https://www.inertiallabs.com/mru-datasheet}, [Online; accessed
  19-July-2021] (2021).

\bibitem{um2017data}
T.~T. Um, F.~M. Pfister, D.~Pichler, S.~Endo, M.~Lang, S.~Hirche, U.~Fietzek,
  D.~Kuli{\'c}, Data augmentation of wearable sensor data for parkinson’s
  disease monitoring using convolutional neural networks, in: Proceedings of
  the 19th ACM International Conference on Multimodal Interaction, 2017, pp.
  216--220.

\bibitem{ohashi2017augmenting}
H.~Ohashi, M.~Al-Nasser, S.~Ahmed, T.~Akiyama, T.~Sato, P.~Nguyen, K.~Nakamura,
  A.~Dengel, Augmenting wearable sensor data with physical constraint for
  {DNN}-based human-action recognition, in: ICML 2017 times series workshop,
  2017, pp. 6--11.

\bibitem{steven2018feature}
O.~Steven~Eyobu, D.~S. Han, Feature representation and data augmentation for
  human activity classification based on wearable imu sensor data using a deep
  lstm neural network, Sensors 18~(9) (2018) 2892.

\end{thebibliography}
\end{document}